\documentclass[aps,pra,letterpaper,twocolumn,showpacs,superscriptaddress,floatfix,longbibliography]{revtex4-1}
\usepackage[english]{babel}
\usepackage{amsmath}
\usepackage{amsfonts}
\usepackage{amssymb}
\usepackage{color}
\usepackage{epsfig}
\usepackage{graphicx}
\usepackage{bm}
\usepackage{mathtools}
\usepackage{bbold}
\usepackage{times}
\usepackage[colorlinks,urlcolor=blue,bookmarks=false,hypertexnames=true]{hyperref} 
\usepackage{flushend}

\usepackage{mathtools}
\usepackage{empheq}
\usepackage{pifont}

\allowdisplaybreaks

\newcommand{\nesssymb}{\infty}

\newcommand{\trb}{\mathop{\mathrm{tr}_{1,N}}\limits}

\definecolor{myred}{RGB}{168,5,14}

\definecolor{myblue}{RGB}{14,5,168}

\newcommand{\xmark}{\hspace{1pt}\ding{55} }

\global\long\def\bra#1{\langle #1 |}
\global\long\def\ket#1{|#1\rangle }

\global\long\def\al{\alpha}
\global\long\def\be{\beta}
\global\long\def\ga{\gamma}
\global\long\def\de{\delta}

\global\long\def\th{\theta}

\global\long\def\la{\lambda}
\global\long\def\ka{\kappa}
\global\long\def\si{\sigma}
\global\long\def\vfi{\varphi}

\newcommand{\traccazero}{\mathop{\mathrm{tr}_{\mathcal{H}_0}}\limits}

\global\long\def\bege{\begin{equation}}
\global\long\def\ende{\end{equation}}

\global\long\def\begal{\begin{align}}
\global\long\def\endal{\end{align}}

\begin{document}

\title{Dissipative generation of pure steady states and a gambler's ruin
  problem }

\author{Vladislav Popkov} \affiliation{Department of Physics,
  University of Wuppertal, Gaussstra\ss e20, 42119 Wuppertal, Germany}
\affiliation{Institut f\"{u}r Theoretische Physik, Universit\"{a}t
  K\"{o}ln, Z\"ulpicher str. 77, K\"{o}ln, Germany}
\affiliation{Faculty of Mathematics and Physics, University of
  Ljubljana, Jadranska 19, SI-1000 Ljubljana, Slovenia}

\author{Simon Essink} \affiliation{HISKP, University of Bonn,
  Nussallee 14-16, 53115 Bonn, Germany}

\author{Corinna Kollath} \affiliation{PI, University of Bonn,
  Nussallee 12, 53115 Bonn, Germany}

\author{Carlo Presilla} \affiliation{Dipartimento di Fisica, Sapienza
  Universit\`a di Roma, Piazzale Aldo Moro 2, Roma 00185, Italy}
\affiliation{Istituto Nazionale di Fisica Nucleare, Sezione di Roma 1,
  Roma 00185, Italy}


\begin{abstract}
  We consider an open quantum system, with dissipation applied
    only to a part of its degrees of freedom, evolving via a quantum
    Markov dynamics.  We demonstrate that, in the Zeno regime of large
    dissipation, the relaxation of the quantum system towards a pure
    quantum state is linked to the evolution of a classical Markov
    process towards a single absorbing state.  The rates of the
    associated classical Markov process are determined by the original
    quantum dynamics.  Extension of this correspondence to absorbing
    states with internal structure allows us to establish a general
    criterion for having a Zeno-limit nonequilibrium stationary state
    of arbitrary finite rank.  An application of this criterion is
    illustrated in the case of an open $XXZ$ spin-1/2 chain dissipatively
    coupled at its edges to baths with fixed and different
    polarizations.  For this system, we find exact nonequilibrium
    steady-state solutions of ranks $1$ and $2$.
\end{abstract}

\maketitle

\section{Introduction}
The dynamics of a classical Markov process with an absorbing state,
e.g., a so-called gambler's ruin problem \cite{DeMoivre}, stops once
the absorbing state is reached, i.e., the gambler has no more coins
left. All the other states in which the gambler has a finite number of
coins, provided that a bank has an infinite money reserve, are
transitory states. Reaching an absorbing state marks the end of the
time evolution.

One of our aims is to point out that, surprisingly, the Markovian
non-unitary evolution of an open quantum system affected by
dissipation towards a pure quantum state can be linked to a classical
Markov process with an absorbing state.  This link is meaningful and
nontrivial if the dissipation acts only on a part of the degrees of
freedom of the quantum system and provided that the dissipation is
strong, i.e., in the so called quantum Zeno
regime~\cite{Misra1977,ZenoMeasurements,ZenoStaticsExperimentalReview}.

Markovian dynamics of an open quantum system is described by a
Lindblad master equation (LME), and part of the degrees of freedom
corresponding to the eigenstates of the system Hamiltonian evolve via
a classical Markov process (MP), the so-called Pauli master
equation~\cite{Petruccione}.  This MP, however, does not provide
substantial information about the nonequilibrium stationary state
(NESS) of the system, since the eigenstates of the system Hamiltonian
do not coincide, generically, with those of the NESS.  The situation
becomes different in the Zeno regime which is governed via an
effective Hamiltonian~\cite{2014Venuti,PhysRevA.94.052339}, commuting
with the Zeno NESS~\cite{2014Venuti}.  The quantum Zeno regime is a
widely used experimental tool nowadays
~\cite{Itano1990,Kwiat1995,Signoles2014,
  Schafer2014,Patil2015,PhysRevLett.97.260402}, and has applications
ranging from engineering dissipative quantum gates~\cite{Beige2000}
and topological states~\cite{1367-2630-15-8-085001} to quantum
simulators~\cite{Stannigel2013} and universal quantum
computations~\cite{Verstraete2009,Yi2012,Elliott2015,
  Winkler2006,PhysRevA.93.022312}.

It has been shown in \cite{2018ZenoDynamics}, that the Lindblad
temporal evolution of the reduced density matrix in the Zeno limit can
be described, at the final stage of relaxation, in terms of an
auxiliary classical Markov process, with rates obtainable from the
original quantum system.  In this auxiliary Markov process, the state
probabilities are the populations of the eigenstates of the
dissipation-projected Hamiltonian of the quantum system (see later),
and converge ultimately to the NESS eigenvalues.  Now, suppose that
the auxiliary Markov process has an absorbing state, i.e., all
populations, except one, vanish in time. Its quantum counterpart, the
quantum density matrix, will relax, in time, to a NESS with only one
eigenstate being populated, i.e., to a pure quantum state.
Conversely, the convergence of a quantum NESS towards a pure state (a
rather exceptional scenario, given that a generic quantum state is
mixed) implies that the corresponding auxiliary Markov process has an
absorbing state.

Exploring further the above analogy is fruitful since we can use the
well-developed theory of classical Markov processes with absorbing
states~\cite{Kemeny} for investigating open quantum systems.  As a
first step, we establish a criterion for a Zeno NESS (assumed unique)
to have an arbitrary rank $r$.  Our criterion contains a classical
part, inherited from the theory of Markov processes with absorbing
states, and an intrinsic quantum part.  The criterion is formulated in
terms of the spectral problem of the dissipation-projected
Hamiltonian, which is drastically simpler then the original
Liouvillian problem.

We start by outlining our setup and revisiting the connection between
the Zeno-limit dynamics of the reduced density matrix and the
associated classical MPs.  After pointing out a link between a pure
NESS (NESS of rank 1) and a MP with an absorbing state, we extend the
analogy to a NESS of arbitrary rank, and formulate our criterion.  The
criterion is then applied to the paradigmatic one-dimensional $XXZ$
spin model with dissipative boundary driving, For this system we find
exact Zeno-limit NESS solutions with rank $r=1$ and $2$.

 \section{Finite rank NESS and its connection to a Markov process}
 We consider an open quantum system with Hilbert space $\mathcal{H}$
 of finite dimension $d$ and a dissipation acting only on a part of
 its degrees of freedom, namely, the subspace $\mathcal{H}_0$ of
 dimension $d_0$.  We assume that the global Hilbert space can be
 partitioned as a tensor product of $\mathcal{H}_0$ and
 $\mathcal{H}_1$, $\mathcal{H} =\mathcal{H}_0 \otimes \mathcal{H}_1$,
 where $\mathcal{H}_1$ is the remaining part of the Hilbert space not
 directly coupled to the dissipation.  The time evolution of the
 system is described via the LME
 \cite{Petruccione,Schaller}
 \begin{align}
   \frac{\partial \rho }{\partial t}
   &=
     -\frac{i}{\hbar} \left[  H,\rho \right] + \Gamma \mathcal{D} [\rho].
     \label{LME}
 \end{align}
 Here the Hamiltonian $H$ describes the unitary part of the dynamics,
 and $\mathcal{D} $ is a Lindbladian dissipator describing the
 interaction with the environment via Lindblad operators $L_k$,
 \begin{align}
   \mathcal{D} [\rho]
   &= \sum_k \mathcal{D}_{L_k} [\rho]
     \nonumber \\
   &= \sum_k \left( L_k \rho L_k^\dagger - \frac{1}{2} L_k^\dagger L_k
     \rho - \frac{1}{2}\rho L_k^\dagger L_k\right).
     \label{DefLMEDissipator}
 \end{align}
 We assume that the effective dissipation strength $\Gamma$ is much
 stronger than the unitary part of the evolution, and that the
 dissipator $\mathcal{D}$ alone targets a unique steady state with the
 density matrix $\psi_0$ in $\mathcal{H}_0$, namely,
 $\mathcal{D} \psi_0=0$.

 In the limit of large $\Gamma$ and time $t\gg 1/\Gamma$, the Lindblad
 dynamics is effectively limited to the decoherence-free subspace,
 namely, $\rho(t)\approx \psi_0\otimes R(t)$.  In fact, it has been
 shown in \cite{2018ZenoDynamics} that, if the dissipator is
 diagonalizable with spectrum $\mathcal{D} \psi_n=\xi_n \psi_n$ ($n$
 labeling different right eigenstates of $\mathcal{D}$), then, for
 $t\gg 1/\Gamma$, $\| \rho(t) - \psi_0\otimes R(t) \| = O(1/\Gamma)$,
 where $R(t)$ is determined by the renormalized master equation
 \begin{align}
   \frac{\partial R}{\partial t} = -\frac{i}{\hbar} [ \tilde H,R ] +
   \frac{1}{\Gamma} \tilde{\mathcal{D}}[R(t)],
   \label{LMEforR}
 \end{align}
 where $\tilde H$ and $\tilde{\mathcal{D}}$ describe effective unitary
 and dissipative temporal evolution within the dissipation-free
 subspace, and can be calculated from $H$ and $\mathcal{D}$ of the
 original LME (\ref{LME}) with the help of the
 Dyson expansion with respect to the $1/\Gamma$ parameter
 \cite{2018ZenoDynamics}. In general, see \cite{2018ZenoDynamics},
 $\tilde H$, apart from the zeroth-order term
 $\lim_{\Gamma \rightarrow \infty} \tilde H =h_D$, can have a
 $O(1/\Gamma)$ correction. In the following we set $\hbar=1$.  The
 dissipation projected Hamiltonian $h_D$, is given by
 ~\cite{PhysRevA.91.052324,2018ZenoDynamics}
 \begin{align}
   h_D
   &= \traccazero \left[ \left(\psi_0 \otimes
     I_{\mathcal{H}_1}\right)H \right],
     \label{h_D}
 \end{align}
 i.e., $h_D$ corresponds to a projection of the original Hamiltonian
 $H$ on the decoherence-free subspace governed by the kernel $\psi_0$
 of the dissipator $\mathcal{D}$.

 For simplicity, we assume $\tilde H=h_D$, and expand the steady-state
 solution of Eq.~(\ref{LMEforR}) $R_\nesssymb(\Gamma)$ in powers of
 $1/\Gamma$,
 \begin{align}
   R_\nesssymb = \sum_{m=0}^\infty \Gamma^{-m}
   R_\nesssymb^{(m)}.
   \label{NessExpansion}
 \end{align}
 The series is convergent for sufficiently large $\Gamma$.

 We are mainly interested in the final nonequilibrium steady state
 described by the density matrix $\psi_0\otimes R_\nesssymb^{(0)}$,
 where
 $R_\nesssymb^{(0)}= \lim_{\Gamma\rightarrow \infty}
 \lim_{t\rightarrow \infty} R(t)$.

 We insert the ansatz (\ref{NessExpansion}) in Eq.~(\ref{LMEforR}) for
 the steady state and compare the orders of $1/\Gamma^k$.  The first two
 orders $k=0,1$ yield
 \begin{align}
   [h_D,R_\nesssymb^{(0)}]
   &=0 \label{CondDiss0term}, \\
   -i[h_D,R_\nesssymb^{(1)}]+ \mathcal{\tilde D} [R_\nesssymb^{(0)}]
   &=0. 
     \label{RecurrenceR}
 \end{align}
 Note that further orders $1/\Gamma^k$ with $k>1$ cannot be trusted
 since Eq.~(\ref{LMEforR}) itself was obtained up to terms of order
 $1/\Gamma$.

 Denote by $\ket{\al}$ a common eigenbasis of $h_D$ and
 $R_\nesssymb^{(0)}$ which is always possible to find since both are
 Hermitian and commute (\ref{CondDiss0term}). We write
 $R_\nesssymb^{(0)}$ in this basis as
 \begin{align}
   R_\nesssymb^{(0)}
   &=\sum_{\al=0}^{d_1-1} \nu_\al^\nesssymb \ket{\al}\bra{\al},
     \label{AssumpR0}
 \end{align}
 where $d_1$ is the dimension of the subspace ${\cal H}_1$.  Let us
 now rewrite Eq.~(\ref{RecurrenceR}) as a set of scalar equations
 using the basis $\ket{\al}$, namely
 \begin{align}
   Q_{\al\al'}
   &\equiv \bra{\al} \mathcal{\tilde D}[ R_\nesssymb^{(0)}]-i
     [h_D,R_\nesssymb^{(1)}] \ket{\al'} = 0,
     \nonumber\\
   &\qquad\qquad \al,\al'=0,\ldots,d_1-1. 
     \label{CondSecularOrder1}
 \end{align}
 We postpone an analysis of the complete set $Q_{\al\al'}=0$ and first
 look at diagonal subset $Q_{\al\al}=0$, for all $\al$. Assuming the
 effective dissipator $\mathcal{\tilde D}$ to be of the canonical form
 $\mathcal{\tilde D}\cdot = \sum_k \ga_k ( {\tilde L}_k \cdot {\tilde
   L}_k^\dagger -\frac{1}{2} \{ \cdot, {\tilde L}_k^\dagger {\tilde
   L}_k \} )$, we obtain, after some algebra, a closed set of
 equations for $\nu_\al^\nesssymb$,
 \begin{align}
   &\sum_{\be\neq \al}^{d_1-1} w_{\be\al} \nu_\be^\nesssymb-
     \nu_\al^\nesssymb \sum_{\be\neq \al}^{d_1-1} w_{\al\be} =0,
     \label{LMEClassicalNESS}
   \\
   &w_{\be \al} = \frac{1}{\Gamma}\sum_k \ga_k \left| \bra{\al} \tilde
     L_k \ket{\be}\right|^2.
     \label{NESS-ClassME2}
 \end{align}
 We recognize in (\ref{LMEClassicalNESS}) a steady master equation of
 a Markov process with transition rates $w_{\al \be}$ between the
 states $\al$ and $\be$.  The factor $ \frac{1}{\Gamma}$ in
 Eq.~(\ref{NESS-ClassME2}) signals that the relaxation towards the
 steady-state slows down with an increased dissipation strength, see
 \cite{2014Venuti, 2018ZenoDynamics}.  The Perron-Frobenius theorem
 guarantees the existence of a solution of
 Eq.~(\ref{LMEClassicalNESS}) with non-negative entries
 $\nu_\al^\nesssymb$, which, according to (\ref{AssumpR0}), have the
 physical meaning of the eigenvalues of the reduced density matrix in
 the Zeno limit $ R_\nesssymb^{(0)}$, obeying the normalization
 condition $\sum_\al \nu_\al^\nesssymb=1$.

 In the following we establish a criterion for a Zeno NESS
 $ R_\nesssymb^{(0)}$ to have a reduced rank $r<d_1$. We start with
 the case of pure NESS corresponding to $r=1$, when
 $R_\nesssymb^{(0)} = \ket{0} \bra{0}$ and generalize it
 afterwards. We remark that this case was already considered in
 \cite{2017PopkovPresillaSchmidt}, but now we revisit it using the
 stochastic interpretation.

 \section{Pure NESS}
 A NESS of the form $R_\nesssymb^{(0)} = \ket{0} \bra{0}$, implies the
 existence of a unique steady state solution of the Markov equation
 (\ref{LMEClassicalNESS}).  This means that for all
 $\beta=0, \dots, d_1-1$, the solution fulfills
 $\nu_\be^{\infty}=\de_{0,\be}$.  In the stochastic Markov picture, a
 unique Markov process steady state of the form
 $\nu_\be^{\infty}=\de_{0,\be}$ can arise if and only if the state
 $\al=0$ is an absorbing state of the Markov process. This means that
 the system cannot escape from state zero, namely,
 \begin{align}
   w_{0\al}=0, \qquad \mbox{for any $\al> 0$},
   \label{DefAbsorbing}
 \end{align}
 while all the other states of the Markov process $\al > 0$ are
 transient. Thus, we establish a link between a pure NESS and an
 auxiliary classical Markov process with an absorbing state.

 \section{ NESS of an arbitrary reduced rank $r$}
 Analogously, we interpret a Zeno NESS of rank $r$ smaller than the
 full rank, $r<d_1$,
 \begin{align}
   R_\nesssymb^{(0)}  = 
   \sum_{z=0}^{r-1}\nu_z^\nesssymb \ket{z}\bra {z} 
   \label{AssumpR0finiteRank}
 \end{align}
 as a consequence of the existence of a closed subset of states in the
 auxiliary classical Markov process with rates $w_{\al \be}$
 satisfying closed set condition $w_{z\al}=0$, for all $z<r$ and
 $\al \geq r$.  At this point, however, one should not forget that our
 original problem is an intrinsic quantum problem, not describable by
 just set of eigenvalues $\nu_z^\nesssymb$. In particular, we have up
 to now neglected the conditions arising from off-diagonal values of
 $Q_{\alpha,\alpha'}$.  What is the full set of conditions which
 guarantee the Zeno NESS to have a reduced rank? We postulate that an
 answer to this question is given by the following criterion:

 \subsection{Criterion}. Let $\ket{\al}$ be an eigenbasis of the
 dissipation-projected Hamiltonian $h_{D}$, $\la_\al$ being the
 respective eigenvalues, and let
 $\mathcal{\tilde D}\cdot = \sum_k \ga_k ( {\tilde L}_k \cdot {\tilde
   L}_k^\dagger -\frac{1}{2} \{ \cdot, {\tilde L}_k^\dagger {\tilde
   L}_k \} )$ be the Lindbladian of the effective dynamics.  A state
 $\rho_\nesssymb = \psi_0 \otimes \sum_{z=0}^{r-1}\nu_z^\nesssymb
 \ket{z}\bra {z} $ is a NESS state of Eq.~(\ref{LME}) in the Zeno
 limit, if and only if the following three conditions are satisfied:\\
 (A) states $z=0,1,\dots,r-1$ form a closed ergodic set in the
 associated Markov process, defined via the transition rates
 $w_{\al \be}= \sum_k \ga_k |\bra{\be} {\tilde L}_k \ket{\al}|^2/\Gamma$;\\
 (B) all the other states $r,r+1,\dots,d_1-1$ are transient;\\
 (C) for any state $\ket{\al}$ such that $\la_\al=\la_z$, where $\al$
 belongs to the transient set and $z$ belongs to the closed set, we
 have
 $\bra{z}\sum_k \ga_k {\tilde L}_k^\dagger {\tilde L}_k \ket{\al} =0$.

 Before proceeding to justify the criterion, we note that conditions
 A and B are rooted in the associated classical Markov process,
 while condition C has an intrinsically quantum origin. Let us note
 that for the Zeno-limit NESS to be a pure state ($r=1$) necessary and
 sufficient conditions were verified by an alternative method in
 \cite{2017PopkovPresillaSchmidt}. The conditions given in
 \cite{2017PopkovPresillaSchmidt} are equivalent to our conditions
 A, B and C for $r=1$ which prove the criterion for $r=1$.

 \subsection{Proof of the necessity of the criterion.}.  In brief,
 conditions A and B, together with the stochasticity property of
 the transition matrix, provide the existence of a unique steady-state
 solution of the associated Markov process, with
 $\nu_0, \nu_1, \dots, \nu_{r-1}$ nonzero and $\nu_k=0$, for
 $k\geq r$.

 To see this, consider the full set of scalar equations in
 (\ref{CondSecularOrder1}).  Noticing that
 $\bra{\al}[h_D,R_\nesssymb^{(1)}]\ket{\al}=0$ and using
 Eq.~(\ref{AssumpR0finiteRank}), we find that the equations
 $Q_{\al\al}=0$ for $\al\geq
 r$ are trivially satisfied. The equations
 $Q_{zz}=0$, with $z=0,1,\dots,
 r-1$ yield a steady-state master equation (\ref{LMEClassicalNESS})
 for steady-state probabilities
 $\nu_{z}^\nesssymb$ inside the closed set.

 Condition C enters the criterion only if the Hamiltonian $h_D$ has
 a special degeneracy, namely, $\la_\al=\la_z$, where $z$ belongs to
 the closed set and $\al$ belongs to the transient one.
 
 The off-diagonal conditions in Eq.~(\ref{CondSecularOrder1}), namely,
 $Q_{z\al}=0$, with $z\leq r-1$ and $\al \geq r$, yield, after some
 algebra,
 \begin{align}
   -i (\la_z-\la_\al) \bra{z} R_\nesssymb^{(1)}\ket{\al}
   &=
     \frac{\nu_z^\nesssymb}{2} \bra{z} \sum_k \ga_k \tilde{L}_k^\dagger
     \tilde{L}_k \ket{\al}.
     \label{CondQzal}
 \end{align}
 In deriving Eq.~(\ref{CondQzal}) we used the properties for
 $\alpha\geq r$ $R_\nesssymb^{(0)}\ket{\al}=0$ and
 $\bra{z} \tilde{L}_n^\dagger \ket{\al}=0$. The latter follows from
 the closed set property A, namely,
 $w_{ z \al}= \sum_n {\ga_n |\bra{\al} \tilde{L}_n \ket{z}}|^2=0$.
 Note that by virtue of our assumption (\ref{AssumpR0finiteRank}) we
 have $\nu_z^\nesssymb \neq 0$, thus, if $\la_z=\la_\al$, the right-hand side of
 Eq.~(\ref{CondQzal}) must vanish, yielding condition C of the
 criterion.  In conclusion, conditions A, B and C are indeed
 necessary for the NESS in the Zeno limit to have rank $r$. $\square$

 \subsection{Argument for sufficiency of the criterion.}  Whereas the
 necessity of the conditions in the criterion is proven, the question
 of sufficiency appears more delicate. For the pure state case, $r=1$,
 sufficiency of our criterion has been rigorously proved in
 \cite{2017PopkovPresillaSchmidt}.  For higher ranks $r>1$ we do not
 have a rigorous argument so far. Instead, here we checked our
 criterion for $r=2$ numerically for a specific model, dissipatively
 driven spin chains.

 \subsection{Reformulation of conditions A and B.}  In the
 following we show how to check conditions A and B given the
 rates $w_{\al \be}$.

 Condition A is equivalent to
 \begin{align}
   w_{z\al}
   &=0, \qquad \mbox{for all $z\leq r-1$ and $\al \geq r$},
     \label{CondClosedSet}
 \end{align}
 complemented with the requirement of ergodicity: each state within
 the closed set is reachable from any other state in the closed set in
 a finite number of steps.

 In order to check condition B, write the transition matrix $F$ (the
 matrix with elements $F_{\be \al}=w_{\al \be}$ for $\be \neq \al$,
 $F_{\al \al}=- \sum_{\be \neq \al} F_{\be \al}$) satisfying the
 closed set conditions (\ref{CondClosedSet}) in the canonical form
 \cite{Kemeny}
 \begin{align}
   F
   &=
     \left(
     \begin{array}{cccccc}
       \mathcal{X}_{00} & \ldots & \mathcal{X}_{0,r-1}
       & \mathcal{R}_{0,r} & \ldots & \mathcal{R}_{0,d_1-1} \\
       \vdots & \vdots & \vdots & \vdots & \vdots & \vdots\\
       \mathcal{X}_{r-1,0} & \ldots & \mathcal{X}_{r-1,r-1}
       & \mathcal{R}_{r-1,r} & \ldots & \mathcal{R}_{r-1,d_1-1}\\
       0 & \ldots & 0
       & \mathcal{K}_{rr} & \ldots & \mathcal{K}_{r,d_1-1}\\
       \vdots & \vdots & \vdots & \vdots & \vdots & \vdots\\
       0 & \ldots & 0
       & \mathcal{K}_{d_1-1,r} & \ldots & \mathcal{K}_{d_1-1,d_1-1}
     \end{array}\right)\nonumber\\
   &=
     \left(
     \begin{array} {cc}
       \mathcal{X} &  \mathcal{R} \\
       0 &   \mathcal{K}\\
     \end{array}
   \right),
   \label{DefTransitionMatrixAbsorbingState}
 \end{align}
 where $ \mathcal{X}_{ii}= - \sum_{j\neq i} \mathcal{X}_{ji} $ and
 $ \mathcal{K}_{ii}= - \sum_{j\neq i} \mathcal{K}_{ji} - \sum_{j}
 \mathcal{R}_{ji}$.  The steady-state equation
 $\sum_{\be=0}^{d-1} F_{\al \be} \nu_\be^\nesssymb=0$ has an obvious
 closed set solution, namely,
 $\nu^\nesssymb = (\nu_0^\nesssymb,\nu_1^\nesssymb,\dots,
 \nu_{r-1}^\nesssymb, 0,0,\dots,0)$, where the nonzero components
 $\nu_z^\nesssymb$ satisfy
 $\sum_{z'=0}^{r-1} \mathcal{X}_{z z'} \nu_{z'}^\nesssymb=0$. This
 solution is unique if and only if all the other states are transient,
 which is equivalent to the absence of zero eigenvalues in the
 spectrum of $\mathcal{K}$, i.e.,
 \begin{align}
   \det \mathcal{K} \neq 0 \label{EqDetK_neq0}.
 \end{align}
 Indeed, assume $\det \mathcal{K} \neq 0$ and split the equations
 $\sum_{\be=0}^{d_1-1} F_{\al \be} \nu_\be^\nesssymb=0$ in two sets,
 \begin{align}
   &\sum_{z'=0}^{r-1} \mathcal{X}_{z z'} \nu_{z'}^\nesssymb
     +\sum_{\al=r}^{d_1-1} \mathcal{R}_{z\al} \nu_{\al}^\nesssymb=0,
     \qquad z=0,1,\dots, r-1,
     \label{EqRnu=0}
   \\
   &\sum_{\be=r}^{d_1-1} \mathcal{K}_{\al \be}
     \nu_{\be}^\nesssymb=0,\qquad \al=r,r+1,\dots,d_1-1.
     \label{EqKnu=0}
 \end{align}
 By the assumption (\ref{EqDetK_neq0}), Eq.~(\ref{EqKnu=0}) has only
 the trivial solution $\nu_{\be}^\nesssymb=0$, for all $\be>0$. Thus
 the closed set solution is the only solution of Eqs.~(\ref{EqRnu=0})
 and (\ref{EqKnu=0}).  Conversely, if $\det \mathcal{K} = 0$, then a
 nontrivial solution of Eq.~(\ref{EqKnu=0}) exists, and, due to the
 Perron-Frobenius theorem, it has non negative real entries. This
 corresponds to the existence of another closed set of states within
 the ``transient'' set of states $\al= r,r+1, \dots,d_1-1$, and,
 therefore, to a violation of the transient hypothesis. Consistency of
 Eq.~(\ref{EqRnu=0}) is guaranteed by the decomposition theorem for
 finite Markov chains~\cite{Kemeny}.

 \section{Example of a dissipatively driven one-dimensional $XXZ$
   spin chain}
 In the following, our findings are illustrated with examples based on
 dissipatively driven one-dimensional $XXZ$ spin chains.

 Consider a chain with $N$ sites occupied by spins $s=1/2$ and
 described by the anisotropic Hamiltonian
 \begin{align}
   H = \sum_{n=1}^{N-1} \vec{\sigma}_n \cdot (J \vec{\sigma}_{n+1}),
   \label {DefXXZHam}
 \end{align}
 where $\vec{\sigma}_n=(\sigma_n^x,\sigma_n^y,\sigma_n^z)$ is the
 vector of the Pauli matrices at site $n$ and
 $J=\mathrm{diag}(J_x,J_y,J_z)$ the anisotropy tensor of the exchange
 interaction.  We choose a dissipation acting at the boundary sites
 $1$ and $N$, and targeting two arbitrary single qubit states
 \begin{align}
   &\rho_{L,R}=\mu_{L,R}  \vec{n}(\th_{L,R},\vfi_{L,R}) \vec{\si}_{1,N}
     \label{rhoLR} \\
   &{\vec n} (\th,\vfi) = (\sin\th \cos\vfi, \sin\th \sin\vfi, \cos\th),
     \label{Def_n}
 \end{align}
 where $\mu_{L,R} \vec{n}_{L,R}$ with
 $\vec{n}_{L,R} \equiv \vec{n}(\th_{L,R},\vfi_{L,R}) $ are the target
 polarizations.  Here, $L(R)$ refers to the leftmost (righmost) site
 of the chain.  Specifically, introducing
 $\vec{n}_{L,R}' = \vec{n} \left(
   \frac{\pi}{2}-\th_{L,R},\vfi_{L,R}+\pi \right) $ and
 $\vec{n}_{L,R}'' = \vec{n} \left(
   \frac{\pi}{2},\vfi_{L,R}+\frac{\pi}{2} \right)$, the above
 dissipation is realized by applying two Lindblad operators at both
 sites $1,N$ of the chain
 \begin{align*}
   L_{1,2}^{\rm{L}}
   &=
     (2 \sqrt{2})^{-1}\sqrt{1\pm \mu_L}
     (\vec{n}_{L}' \mp i \vec{n}_{L}'')\vec{\si}_{1}
   \\
   L_{1,2}^{\rm{R}}
   &=
     (2 \sqrt{2})^{-1}\sqrt{1\pm \mu_R}
     (\vec{n}_{R}' \mp i \vec{n}_{R}'')\vec{\si}_{N}.
 \end{align*}
 A realization of the Lindblad dynamics~(\ref{LME}) for the present
 class of models can for example be obtained via a protocol of
 repeated interactions: the edge spins of the lattice are brought into
 interaction with two ``bath'' qubits, the latter qubits being kept in
 fixed, mixed, or pure states $\rho_L$ and $\rho_R$ from
 (\ref{rhoLR}); the interaction has strength $\gamma_{B}$ and repeats
 periodically at time intervals $\tau$ each time with a bath qubit in
 the fixed state. Within this protocol, the time evolution from time
 $t$ to time $t+\tau$ is described by the map
 \begin{align}
   \rho_{t+\tau} = \mathop{\mathrm{tr}_{0,N+1}}\limits \left( e^{-i
   {\cal{H}} \tau} \rho_L \otimes \rho_{t} \otimes \rho_R e^{i
   \cal{H} \tau} \right) ,
   \nonumber
 \end{align}
 where sites $0,N+1$ denote the qubits from the left and right baths,
 respectively, and ${\cal{H}} =H + \gamma_{B} (U_{0,1} + U_{N,N+1})$
 is the Hamiltonian of the original system plus the system-bath
 interactions $ U_{0,1}$ and $U_{N,N+1}$. The LME dynamics (\ref{LME})
 follows from the above discrete map in the limit
 $\tau \rightarrow 0$, $\gamma_B^2 \tau \rightarrow \Gamma$ (see
 \cite{Karevski}).

 In \cite{2018ZenoDynamics} we have demonstrated that in the Zeno
 limit the effective dynamics of the spin model is described by a LME
 of type (\ref{LMEforR}) for $R(t)=\trb \rho(t)$, with
 ${\tilde H}=h_D$ and
 $\mathcal{\tilde D}=\mathcal{\tilde D}_L+\mathcal{\tilde D}_R$ given
 by
 \begin{align*}
   &h_D = \sum_{j=1}^{M-1} \vec{\sigma}_j \cdot (J \vec{\sigma}_{j+1})
     +\mu_L (J\vec{n}_L) \cdot \vec{\sigma}_1 + \mu_R (J\vec{n}_R) \cdot
     \vec{\sigma}_M,
   \\
   &\mathcal{\tilde D}_{L} = 2(1+\mu_{L}) D_{g_{1L}} + 2(1-\mu_{L})
     D_{g_{1L}^\dagger} + \frac{1} {2}(1-\mu_{L}^2) D_{g_{3L}},
   \\
   &\mathcal{\tilde D}_{R} = 2(1+\mu_{R}) D_{g_{1R}} + 2(1-\mu_{R})
     D_{g_{1R}^\dagger} + \frac{1} {2}(1-\mu_{R}^2) D_{g_{3R}}.
 \end{align*}
 Here the operators act in ${\cal H}_1$, the Hilbert space of the
 inner spins $2,3,\dots,N-1$. We renumerate the inner spins as
 $1,\dots,M=N-2$.  The operators $D_{g}$ are Lindblad dissipators of
 standard form defined in terms of the effective Lindblad operators
 $g_{1L} = (J\vec{n}_{L}') \cdot \vec{\sigma}_1 - i (J\vec{n}_L'')
 \cdot \vec{\sigma}_1$,
 $g_{3L} = 2 (J\vec{n}_L) \cdot \vec{\sigma}_1$,
 $g_{1R} = (J\vec{n}_{R}') \cdot \vec{\sigma}_M - i (J\vec{n}_R'')
 \cdot \vec{\sigma}_M$ and
 $g_{3R} = 2 (J\vec{n}_R) \cdot \vec{\sigma}_M$.
 
 \section{Pure state $r=1$}
 As a first example of application of our finite rank Zeno NESS
 criterion, we consider a NESS being a pure spin-helix state (SHS)
 $\rho_\nesssymb=
 \ket{\xi} \bra{\xi}$, where
 \begin{align}
   \ket{\xi}
   &=
     \bigotimes_{k=1}^{N} \left(
     \begin{array}{c}
       \cos(\frac{\theta}{2})  e^{-\frac{i}{2} \varphi (k-1)}
       \\
       \sin(\frac{\theta}{2})  e^{\frac{i}{2} \varphi (k-1)}
     \end{array}
   \right).
   \label{SHS}
 \end{align}
 This state describes a frozen wave-like spin structure, formed by a
 rotation of a local spin around the $z$-axis along the chain, with
 constant azimuthal angle difference $\varphi$ between neighboring
 spins. Indeed, the expectation value of the local spin at site $k$ is
 \begin{align*}
   \bra{\xi}\vec{\sigma}_k \ket{\xi} = (\sin\theta \cos [\varphi
   (k-1)],\sin\theta \sin [\varphi (k-1)],\cos\theta).
 \end{align*}
 
 We rewrite the factorized NESS (\ref{SHS}) in the form evidencing
 left and right dissipation target states, namely,
 $\rho_\nesssymb = \psi_0^L \otimes \ket{0}\bra {0} \otimes \psi_0^R$.
 This corresponds to choosing $\mu_L=\mu_R=1$, $\th_L=\th_R=\th$,
 $\varphi_L=0, \varphi_R=\vfi(N-1)$ in Eq.~(\ref{rhoLR}), and
 \begin{align}
   \ket{0}&=
            \bigotimes_{k=1}^{M} \left(
            \begin{array}{c}
              \cos(\frac{\theta}{2})  e^{-\frac{i}{2} \varphi k}
              \\
              \sin(\frac{\theta}{2})  e^{\frac{i}{2} \varphi k}
            \end{array}
   \right).
   \label{Defshs}
 \end{align}

 For the given values of $\mu_L$ and $\mu_{R}$ the original
 Lindbladian dissipator has two nonzero Lindblad operators $L^L_1 $
 and $L^R_1$, and the target boundary states are pure (target mixed
 boundary states would typically lead to a full rank Zeno NESS).  The
 corresponding effective Lindblad operators ${\tilde L}_1 = g_{1L}$
 and ${\tilde L}_2 = g_{1R}$.  The associated effective dissipator is
 \begin{align}
   \mathcal{\tilde D}&= 4 \mathcal{ D}_{g_{1L}} + 4 \mathcal{
                       D}_{g_{1R}}.
                       \label{DefTildeDpure}
 \end{align}

 In \cite{2016PopkovPresilla,2017PopkovSchmidtPresilla} it has been
 shown that Eq.~(\ref{SHS}) is the exact Zeno-limit NESS of the
 boundary driven $XXZ$ spin chain with anisotropy
 $J_x=J_y=J_z/\Delta$, provided $\Delta =\cos \varphi$, targeted
 single spin states $\psi_0^{L,R}$ fit the SHS (\ref{SHS}), and
 (\ref{SHS}) does not contain collinear spins.

 We now turn to discuss the connection to the Markov process. The
 stochastic transition matrix of the auxiliary Markov process is
 defined via the rates $w_{\al\be}$ given by
 Eq.~(\ref{NESS-ClassME2}), where $\ket{\al}$ and $\ket{\be}$ are the
 eigenvectors of $h_D$. We thus find
 \begin{align}
   w_{\al\be}&= \frac{4}{\Gamma} \left( | \bra{\be} g_{1L} \ket{\al}
               |^2 + | \bra{\be} g_{1R} \ket{\al} |^2 \right).
               \label{DefwForpure}
 \end{align}

 One can check that (\ref{Defshs}) is an eigenstate of $h_D$ [with
 eigenvalue $\la_0=(N-1)J \cos \vfi$], and of $g_{1R},g_{1L}$ with
 eigenvalues $\pm \kappa$, $\ka = i J \sin \th \sin \vfi$.  Using the
 orthogonality of the eigenbasis of $h_D$, we get
 $\bra{\al} g_{1L} \ket{0} = \bra{\al} g_{1R} \ket{0} = 0$ for all
 $\al > 0$, leading [see Eq.~(\ref{DefwForpure})] to the absorbing
 state condition (\ref{DefAbsorbing}). The relaxation of the global
 reduced density matrix towards a pure NESS thus corresponds, in the
 language of the auxiliary Markov process, to a convergence of the
 classical Markov process towards an absorbing state.  Furthermore, it
 has been proven in \cite{2017PopkovPresillaSchmidt} via a different
 method that in the Zeno limit the NESS converges to the pure state
 $\ket{\xi}\bra{\xi}$ if and only if Eq.~(\ref{EqDetK_neq0}) and
 condition C are satisfied, in accordance with the criterion for
 $r=1$.
 
 \section{NESS of rank $r=2$}
 Here we apply our criterion to find a Zeno-limit NESS of rank $2$, in
 the same model.  We can set the parameters in such a way that
 condition A for $r=2$ is satisfied, namely, there exists a closed set
 of two states in the associated Markov process. Choose the first and
 the last spin of the chain to be dissipatively projected into
 parallel or antiparallel states, $\vfi_R-\vfi_L = n \pi$, while
 keeping $\th_L=\th_R=\th$, and set
 $\Delta=\cos \vfi= \cos (2 m \pi /(N-1)) $ for parallel orientation
 and $\Delta=\cos \vfi= \cos ((\pi +2 m \pi) /(N-1)) $ for
 antiparallel orientation.  Then, two eigenstates of the
 dissipation-projected Hamiltonian can be found, namely, the state
 $\ket{0_+}$ given by Eq.~(\ref{Defshs}) and the state $\ket{0_-}$
 obtained from $\ket{0_+}$ by changing the sign of the helicity,
 $\vfi\to-\vfi$.  Both eigenstates $\ket{0_\pm}$ have the same
 eigenvalue $\la_0 = (N-1)J\cos \vfi $.

 The states $\ket{0_\pm}$ are in general not orthogonal,
 $\langle 0_- | 0_+ \rangle= \eta$, where $\eta$ can be real or
 imaginary, $\eta^*=\pm \eta$, depending on the parameters. The
 overlap $\eta$ can be expressed via a $q$-Pochhammer symbol,
 \begin{align}
   \label{etaGeneral}
   &\eta(\th,\vfi,N)
     = \prod_{k=1}^{N-2} \left(\cos(k\varphi) -i \cos (\theta )
     \sin(k\varphi)\right)
     \nonumber \\
   &\quad= e^{-i\frac{M(M+1)}{2}\vfi} \cos^{2M}\left(\frac{\theta }{2}\right)
     \left(-\tan^2\frac{\theta}{2}; \ e^{2i\varphi} \right)_M .
 \end{align}

 Out of the states $\ket{0_\pm}$, we build the orthonormal
 combinations
 $\ket{0} = 1/\eta_+ [\ket{0_+} + (\eta/|\eta|) \ket{0_-}]$,
 $\ket{1} = 1/\eta_- [\ket{0_+} - (\eta/|\eta|) \ket{0_-}]$, where
 $\eta_\pm=\sqrt{2\pm 2|\eta|}$.  Using the fact that $\ket{0_+}$ is
 an eigenstate of $g_{1R}$ and $g_{1L}$ with eigenvalue $\kappa$ and
 $-\kappa$, respectively, we find
 \begin{align}
   g_{1R} \ket{0}
   &=
     -g_{1L} \ket{0} = a_0 \ket{1}, \qquad a_0 = \kappa
     \eta_{-}/\eta_{+},
   \\
   g_{1R} \ket{1}
   &=-g_{1L} \ket{1} = a_1 \ket{0}, \qquad a_1 = \kappa
     \eta_{+}/\eta_{-}.
     \label{EqNonDiagAction}
 \end{align}
 It follows that the rates of the associated effective Markov process
 are
 \begin{align}
   w_{0\al}&= w_{1\al}=0, \qquad \al \geq 2,
             \label{NESSrank2}\\
   w_{01}&= \frac{8}{\Gamma}|a_0|^2,\\
   w_{10}&= \frac{8}{\Gamma}|a_1|^2.
 \end{align}
     
 Equation~(\ref{NESSrank2}) corresponds to the closed set conditions
 (\ref{CondClosedSet}) for the states with labels $z=0,1$, while
 $w_{10}w_{0,1} \neq 0$ provides the ergodicity of the closed set.
 Moreover, the system $F \nu^{\infty}=0$, equivalent to
 $\nu_0^\nesssymb w_{01}= \nu_1^\nesssymb w_{10}$, is solved by
 $\nu_0^\nesssymb/\nu_1^\nesssymb = w_{10}/w_{01}=
 (1+|\eta|)^2/(1-|\eta|)^2$ and
 $\nu_2^\nesssymb=\nu_3^\nesssymb=\dots=0$.

 The explicit form of the rank $2$ state, that appears in the Zeno
 limit is
 \begin{align}
   \rho_\nesssymb = \psi_0^L \otimes \left(
   \frac{(1+|\eta|)^2}{2+2|\eta|^2} \ket{0}\bra{0} +
   \frac{(1-|\eta|)^2}{2+2|\eta|^2} \ket{1}\bra{1} \right) \otimes
   \psi_0^R
   \label{ResultNESSrank2},
 \end{align}
 with $\psi_0^{L,R} = \rho_{L,R}$.  The topological aspects of the
 state (\ref{ResultNESSrank2}) are discussed in
 \cite{2019CorinnaTopological}.  Let us note that for $\th=\pi/2$ and
 all cases in which this state is the NESS (see below) the overlap
 $\eta$ simplifies to
 \begin{align}
   \eta = \left\{
   \begin{array}{ll}
     2^{2-N},
     &\mbox{if $\vfi_R-\vfi_L=0$ and $N$ even},
     \\
     -(i/2)^{N-2},
     &\mbox{if $\vfi_R-\vfi_L=\pm \pi$ and $N$ even}.
   \end{array}
       \right.
       \label{RESoverlap}
 \end{align}

 Finally, we need to check conditions B and C. Unlike condition A, we
 check them numerically, diagonalizing $h_D$. The results for NESS
 ranks obtained for chains with size $3\leq N \leq 13$ are summarized
 in Table~\ref{tor}. Firstly, we notice that the NESS has rank $r=2$
 if and only if conditions A-C are satisfied, numerically supporting
 the validity of our criterion.

 In addition, analyzing Table~\ref{tor} we notice that a NESS with
 rank $r=2$ occurs if and only if
 \begin{align}
   &\vfi= (\pi m)/(N-1)\label{CondRank2-1}
   \\
   &\mbox{and $N-1$ is not a multiple of $m$}.
     \label{CondRank2-2}
 \end{align}
 This pattern has simple geometrical interpretation, namely, the
 states (\ref{SHS}), which define (\ref{ResultNESSrank2}), do not
 contain any pairs of collinear spins, except for the two boundary
 spins, which are parallel by construction.  E.g. for $N=9$ a rank $2$
 state only appears in the Zeno limit when $\Delta=\cos[ (2 \pi m)/8]$
 with $m=1,3,5,7$.  If $N-1$ is a prime number, (\ref{CondRank2-2}) is
 satisfied for all $\Delta=\cos[ (2 \pi m)/(N-1)]$ $m=1,2,\dots, N-2$.
 \squeezetable
 \begin{table}
   \centering
   \caption{Table of ranks for parallel and antiparallel boundary
     spins computed from the stochastic matrix $F_{\al\be}$ for
     different system sizes $N$.  Note that only $\vfi<\pi$ are
     considered because of the NESS symmetry $\vfi\to-\vfi$.  We also
     show the degeneracy of the eigenvalue $\la_0=(N-1)J \Delta$ of
     $h_D$. The last three columns check conditions A, B and C of the
     criterion for rank $2$ states.  Symbols \checkmark, \xmark, N/A
     and ``$-$'' indicate, respectively, that the corresponding property
     is satisfied, violated, nonapplicable, and noncheckable.
     Property C should only be checked in case of extra degeneracy of
     $\la_0 >2$.  For $N\geq 13$ the NESS rank (whenever it is larger
     than $2$) cannot be determined reliably because of numerical
     precision-related issues \cite{SimonMasterThesis2018} }
   \label{tor}
   \begin{tabular}{cccccccc}
     \hline\hline
     $N$  & $\vfi$                                                                                       & NESS rank & Full rank &  deg($\lambda_0$)  & A & B                                                                                                                                           & C                                                                \\ \hline
     3  & $\frac{\pi}{2}$                                                                                & 2    & 2         & 2             & \checkmark                                        & N/A                                                                                               & N/A                                                                                             \\ 
     4  & $\frac{\pi}{3}, \frac{2\pi}{3}$                                                                & 2    & 4          & 2            & \checkmark                                        & \checkmark                                                                                     & N/A                                                                                             \\ 
     5  & $\frac{\pi}{2}$                                                                                & 8    & 8         & 4             & \checkmark                                        & \xmark                                                                                         &         \checkmark                                             \\
          & $\frac{\pi}{4}, \frac{3\pi}{4}$                                                                & 2    &           & 2             & \checkmark                                        & \checkmark                                                                                     & N/A                                                                                             \\ 
     6  & $\frac{\pi}{5}, ... , \frac{4\pi}{5}$                                                          & 2    & 16          & 2           & \checkmark                                        & \checkmark                                                                                     & N/A                                                                                             \\ 
     7  & $\frac{\pi}{2}$                                                                                & 32   & 32        & 8             & \checkmark                                        &  \xmark                                                                                        &                \checkmark                                                \\
          & $\frac{\pi}{3}, \frac{2\pi}{3}$                                                                & 22   &           & 4             & \checkmark                                        & \checkmark                                                                    		    & \xmark \\
          & $\frac{\pi}{6}, ... , \frac{5\pi}{6}$                                                          & 2    &           & 2             & \checkmark                                        & \checkmark                                                                                     & N/A                                                                                             \\ 
     8  & $\frac{\pi}{7}, ... , \frac{6\pi}{7}$                                                          & 2    & 64          & 2           & \checkmark                                        & \checkmark                                                                                     & N/A                                                                                             \\ 
     9  & $\frac{\pi}{2}$                                                                                & 128  & 128       & 16            & \checkmark                                        &  \xmark                                                                                        &         \checkmark                                                    \\
          & $\frac{\pi}{4}, \frac{3\pi}{4}$                                                                & 72   &           & 4             & \checkmark                                        &  \xmark                                                                                        &            \checkmark                                                     \\
          & $\frac{\pi}{8}, \frac{3\pi}{8}, \frac{5\pi}{8}, \frac{7\pi}{8}$                                & 2    &           & 2             & \checkmark                                        & \checkmark                                                                                     & N/A                                                                                             \\ 
     10 & $\frac{\pi}{3}, \frac{2\pi}{3}$                                                                & 170  & 256       & 8             & \checkmark                                        & \checkmark											    & \xmark                                 \\
          & $\frac{\pi}{9},\frac{2\pi}{9}, \frac{4\pi}{9}, \frac{5\pi}{9}, \frac{7\pi}{9}, \frac{8\pi}{9}$ & 2    &           & 2             & \checkmark                                        & \checkmark                                                                                     &       N/A                                                                                        \\ 
     11 & $\frac{\pi}{2}$                                                                                & 512  & 512        & 32             & \checkmark                                        &  \xmark                                                                                        &                \checkmark                                                \\
          & $\frac{\pi}{5}, \frac{2\pi}{5}$                                                                & 254    &        & 4            & \checkmark                                        & \checkmark											    & \xmark                                 \\
          & $\frac{\pi}{10},\frac{3\pi}{10}, \frac{7\pi}{10}, \frac{9\pi}{10}$				    & 2    &           & 2             & \checkmark                                        & \checkmark                                                                                     &       N/A                                                                                        \\ 
     12 & $\frac{\pi}{11},\dots, \frac{10\pi}{11}$                                                       & 2    & 1024       & 2             & \checkmark                                        & \checkmark											    & N/A                                \\ 
     13 & $\frac{\pi}{2}$                                                                                & 2048  & 2048        & $-$             & $-$                                        &  $-$                                                                                        &  $-$                                               \\
          & $\frac{\pi}{3}, \frac{2\pi}{3}$                                                                                & $-$    &        & 16            & \checkmark                                        & \checkmark											    & \xmark                                 \\
          & $\frac{\pi}{4}, \frac{3\pi}{4}$                                                                & $-$    &        & 8           & \checkmark                                        & \xmark											    & \checkmark                                 \\
          & $\frac{\pi}{6}, \frac{5\pi}{6}$                                                                & $-$    &        & 4            & \checkmark                                        & \xmark											    & \checkmark                                 \\
          & $\frac{\pi}{12},\frac{5\pi}{12}, \frac{7\pi}{12}, \frac{11\pi}{12}$		            & 2    &           & 2             & \checkmark                                        & \checkmark                                                                                     &       N/A                                                                                        \\ \hline\hline

   \end{tabular}
 \end{table}

 \section{Conclusions}
 We have established a link between a quantum dissipative NESS with
 reduced rank and an auxiliary classical Markov process with absorbing
 states (closed sets).  This link paves the way for studies of quantum
 master equations using the well developed theory of classical Markov
 processes. In the present paper, using this link, we suggested a
 criterion for a NESS in the Zeno limit to have a reduced rank.  The
 criterion is illustrated with an example in which rank $1$ (pure
 NESS) and rank $2$ NESS solutions appear in dissipatively boundary
 driven $XXZ$ spin chains .  Noteworthy, our criterion has a
 ``classical Markov process'' part, (properties A and B), and an
 intrinsic quantum part (property C), which has no classical analog,
 being related to a degeneracy of a special eigenvalue in the spectrum
 of the associated dissipation-projected Hamiltonian.  Deep
 understanding of the quantum part of the criterion remains a
 challenge for the future.  From the applicative viewpoint, our
 criterion allows dissipative targeting of pure states or simple
 mixtures of few quantum states, a task of fundamental importance in
 the initialization of quantum simulators~\cite{ini} (see
 also~\cite{TCCG} for an up-to-date review of recent and ongoing
 experiments).

\begin{acknowledgments}
  VP thanks the Department of Physics of Sapienza University of Rome
  for hospitality and financial support. Financial support from the
  Deutsche Forschungsgemeinschaft through DFG projects KL 645/20-1 and
  KO 4771/3-1, projects 277625399 - TRR 185 (B4) and 277146847 - CRC
  1238 (C05) and under Germany's Excellence Strategy – Cluster of
  Excellence Matter and Light for Quantum Computing (ML4Q) EXC 2004/1
  – 390534769 and from the European Research Council (ERC) under the
  Horizon 2020 research and innovation programme, grant agreement
  No.~648166 (Phonton) and No.~694544 (OMNES), and from
  interdisciplinary UoC Forum "Classical and quantum dynamics of
  interacting particle systems" is gratefully acknowledged. We thank
  G. Sch\"utz for critical remarks on the paper.
\end{acknowledgments}



\end{document}